\documentclass[prl,showpacs, twocolumn,superscriptaddress]{revtex4}
\usepackage{graphicx}

\begin{document}

\title
{Magnetism and structure at a vacancy in graphene}

\author{M. W. C. Dharma-wardana}
\email{chandre.dharma-wardana@nrc-cnrc.gc.ca}
\affiliation{
Institute for Microstructural Sciences, National Research Council of Canada}
\author{Marek Z. Zgierski}
\email[Email address:\ ]{marek.zgierski@nrc-cnrc.gc.ca}
\affiliation{Stacie Institute of Molecular Sciences, National Research Council of Canada,
Ottawa, Canada K1A 0R6}

\date{\today}
\begin{abstract}
The electronic structure, bonding and magnetism in 
graphene containing vacancies are studied using density-functional
methods.   
The single-vacancy graphene ground state is
spin polarized and structurally flat.
The unpolarized state is non planar only for finite segments.
Systems containing periodic arrays of vacancies
displays magnetic transitions and
metal-insulator transitions.
\end{abstract}
%
\maketitle
%
{\it Introduction--}
Graphene is a two-dimensional (2D) sheet of carbon atoms
forming a honey-comb lattice, with promising
novel technologies and new 
physics\cite{geim,zhang}. 
These include applications based on carbon
nanotubes\cite{dress,cdw} which are topologically equivalent to
folded
graphene ribbons.
Graphene nanoribbons have
delocalized states only in one dimension, with conductive
and localized edge states.
The localized states appear as ``flat bands''
near the Fermi energy, within tight-binding models which do not treat the lattice
relaxation of the edge atoms\cite{fujita,fertig}.
%

When lattice relaxation is included via first-principles calculations, 
with the dangling bonds saturated with H atoms\cite{louie},
 the electronic structure is found to depend specifically on the ribbon width and 
the relaxed edge structure. Band gaps are found to appear at the Fermi energy.
The role of edges in graphene nanoribbons in producing magnetic ground
states has also been studied using such methods\cite{barone}.

At finite temperatures, entropy as 
well as the 2D structure of graphene favours disorder.
These lead to a quasi-3D wavy conformation\cite{mgknbr,jop}.
Many authors have also used vacancy or defect models\cite{pgcnv}
to include disorder effects in the graphene system. However, our
calculations\cite{grp2} for vacancies and also N-substituted defects
in graphene showed that the energy costs of such defects are 
large enough to prevent natural occurences of such lattice defects, 
thus differing from  metals
or semiconductors. Hence in ref.~\onlinecite{grp2} we concluded that
the graphene samples used in initial quantum Hall-effect studies
were {\it good samples} basically free of vacancies.
However, vacancies and defects can be artificially introduced, and
used to control the energy and spin states near the 
Fermi energy. The vacancies may be positioned lithographically
to generate  graphene containing a periodic
array of vacancies. Such arrays are metastable, and can be annealed to give
various forms of {\it tattered graphene}. The tattering occurs because
two second-neighbour vacancies are found to coalesce into a more stable 
divacancy, which form a sink for other second-neighbour vacancies.
More distant vacancies simply distort the $\sigma$-bonding skeleton
and stabilize themselves. Individual vacancies carry a spin polarization,
and provide a new type of qubit for quantum computational devices. The
variations in spin polarization and band-gaps with the vacancy concentration
set the stage for magnetic and metal-insulator transitions.
Hence here we study an infinite graphene sheet with periodically
distributed vacancies using a plane-wave density-function theory (DFT)
approach based on the Vienna ab-initio simulation code (VASP\cite{vasp}).
We also study finite segments of graphene, with real-space
DFT-calculations using Gaussian basis sets (Gaussian-98 code\cite{gaussian98}). The 
finite graphene studied here 
have zig-zag edges terminated with hydrogen atoms.

\begin{figure} 
\includegraphics*[width=7.0cm,height=7.0cm]{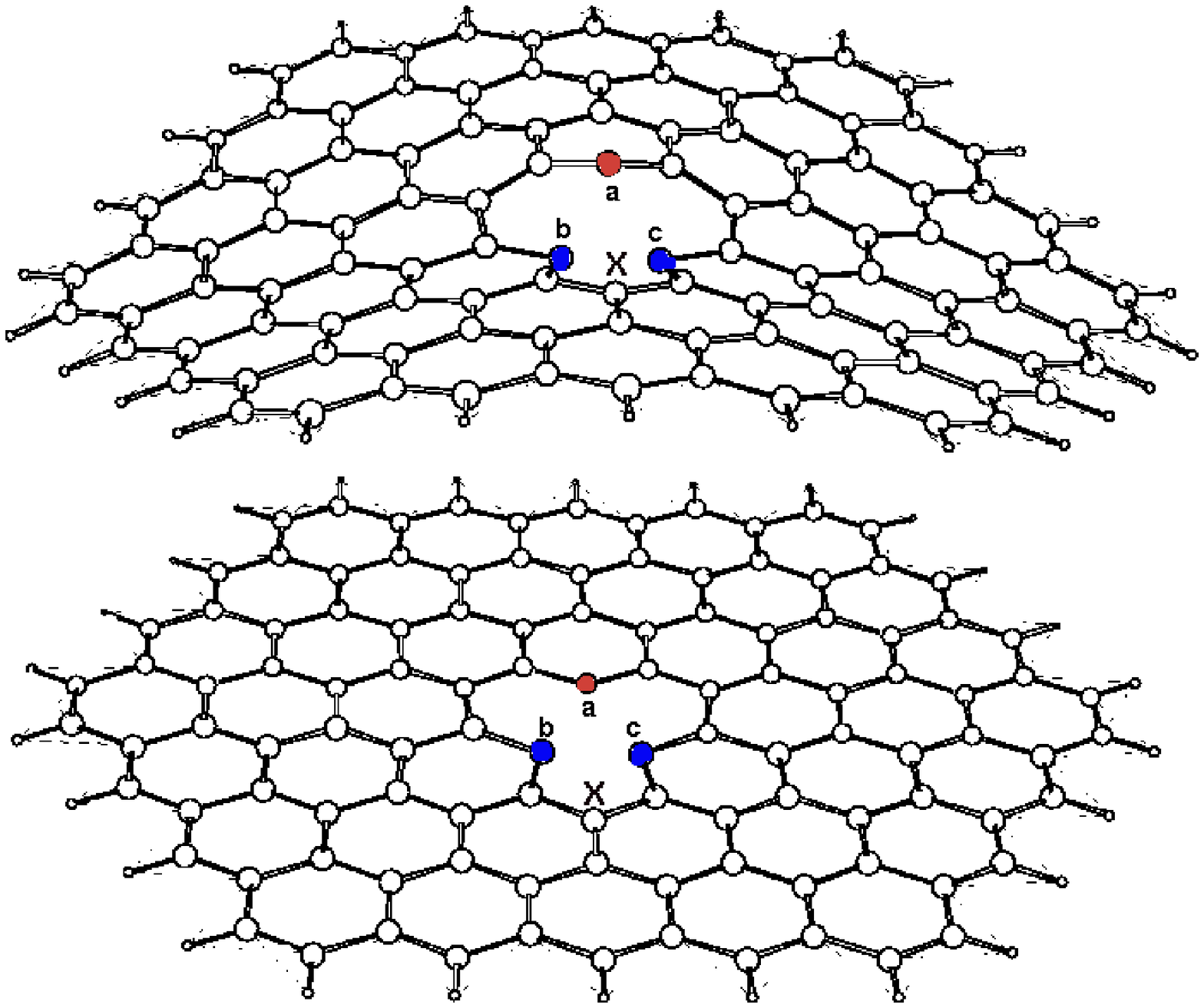}
\hfill
\hfill
\includegraphics*[width=7.0cm,height=4.0cm]{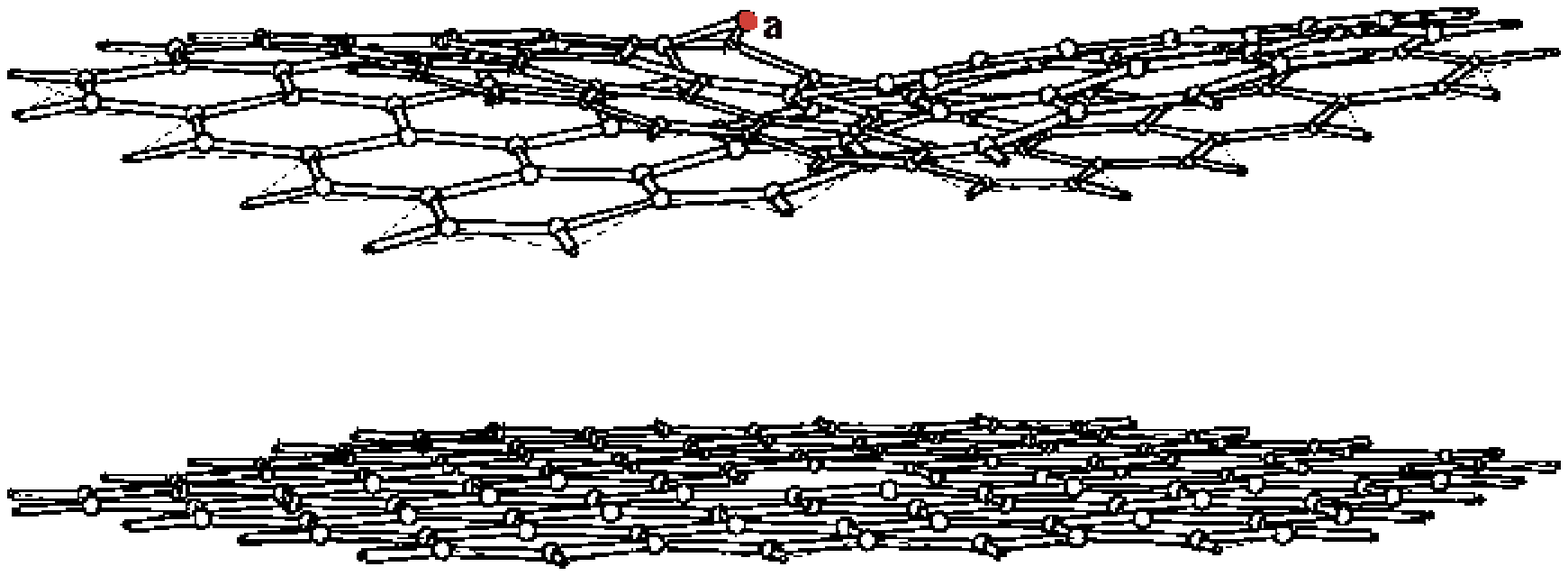}
\caption
{(Color online)Relaxation of a finite sheet (44 hexagons) of graphene containing a vacancy. The
carbon atoms marked ``a, b, c'' were bound to the C atom removed to form the vacancy.
The bent structure shows the non-planar graphene sheet
where the four electrons released by the vacancy
are accomodated to form a spin-unpolarized structure. The ground state
is planar and spin-polarized. Two views of it are shown. The atoms marked
`b', `c' (in blue) are close enough in both structures to form a singlet interaction.
The atom `a' (in red) juts upwards in the unpolarized structure and carries an
unpaired electron. The atoms `X' and `a' have two short bonds in the pentagons.}
\label{mz-both2}
\end{figure}

The finite graphenes with a vacancy give a planar spin-polarized ground state,
while the unpolarized state is slightly higher in energy ($\sim 200$ meV) 
and show a non-planar, saddle-like structure (see Fig.~\ref{mz-both2}).

The creation of the vacancy removes 4 electrons, and
at the same time releases three $sp^2$ electrons (dangling bonds) and a
$\pi$-electron. The $sp^2$-type electrons are associated with
the atoms marked ``a, b, c'' in Fig.~\ref{mz-both2}. The
$\pi$-electron is common to these three sites.
The system distorts or spin polarizes
to accomodate these dangling bonds and the free charges (4 per vacancy).
The vacancy is surrounded by
three pentagons. The {\it tri-pentagonal} structure has one pentagon
smaller than the other two, enabling a singlet
interaction between the atoms ``b'' and ``c'' which are closely positioned,
compared to the pairs ``a, b'' and ``a, c'' of the other two pentagons.
The atom ``a'' in the unpolarized
structure (UPS),
juts upwards (two views of the structure are given in Fig.~\ref{mz-both2}),
 while the atoms 
marked ``b, c'' dip downwards. This distortion disconnects the $\pi$-electron 
generated by the vacancy from
the rest of the 2D electron network. The atom ``a'' of the UPS carries a lone-pair
singlet of electrons, making the total
structure unpolarized. The sheet around the vacancy
acquires a saddle-like distortion. The two bonds associated with
`X', and the atom `a' defining the edge of the vacancy
 are shortened ($\sim 1.39$\AA )
compared to the normal 1.42 \AA $\,$ bond length in graphene.

\begin{figure} 
\includegraphics*[width=6.0cm,height=3.5cm]{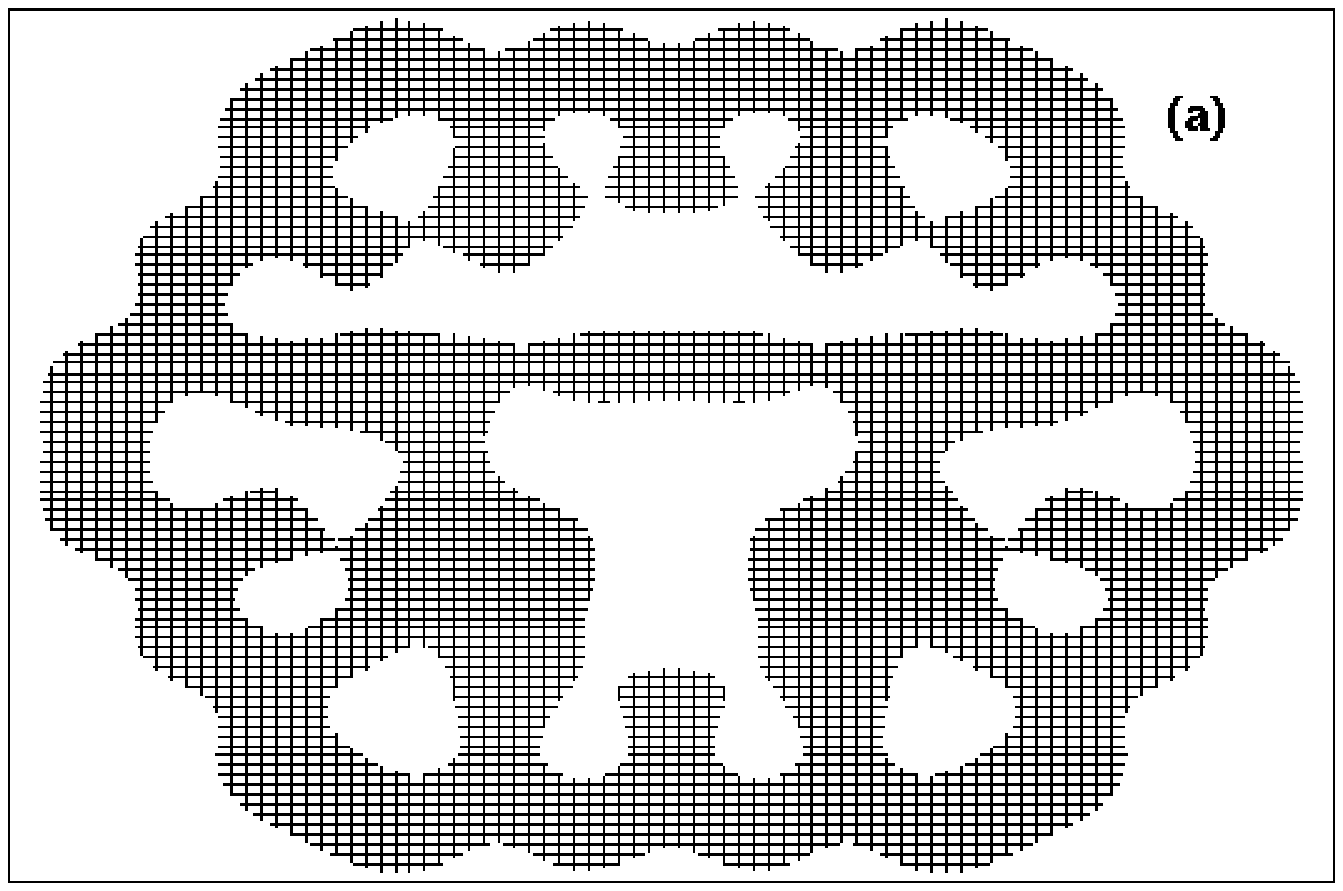}
\hfill
\includegraphics*[width=6.0cm,height=3.5cm]{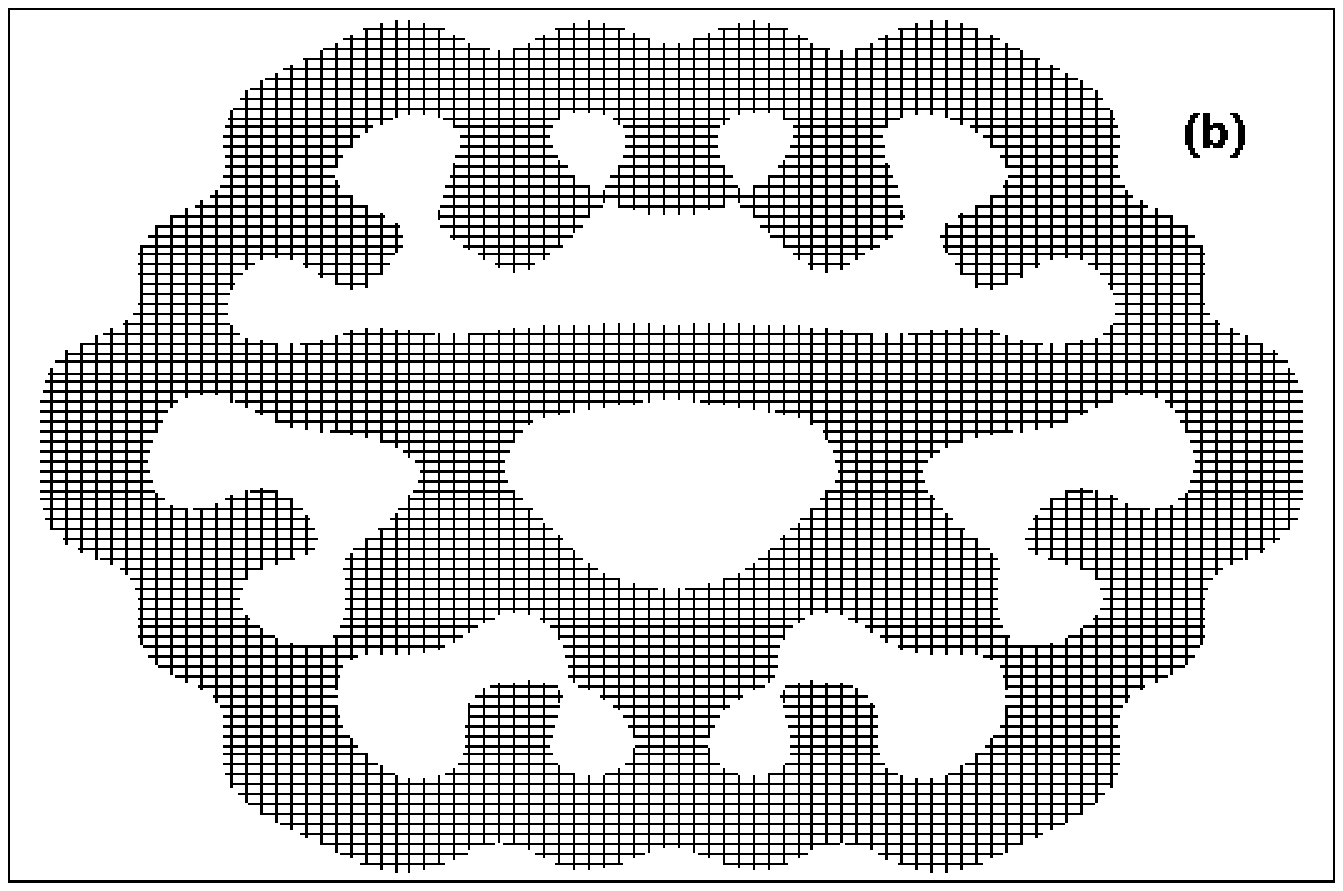}
\caption
{ The distribution of excess charge density (hatched areas) in the 2D plane
of the graphene sheet around a vacancy, for the structure shown in 
Fig.~\ref{mz-both2}, for the spin-unpolarized (a) and polarized (b)
case.}  
\label{charge-print}
\end{figure}

The lower-energy (ground) state is the planar spin-polarized structure (PSPS).
Two views of it are given in Fig.~\ref{mz-both2}.
The stabilization energy for this
system (44 hexagons with a zig-zag edge saturated with H atoms), compared to the
unpolarized state is 217 meV, 
i.e., comparable to room-temperature energies.
This result is obtained with a Gaussian-98 DFT calculation using the
Becke-Lee-Yang-Parr (B3LYP) exchange-
correlation metafunctional\cite{b3lyp}
within a Gaussian (6-31G*, see Ref.~[\onlinecite{acronyms}] for definitions)
 basis set.
The short distance between atoms ``b, c'' enables the two $sp^2$ dangling bonds
to form a singlet. The atom ``a'' now carries one $sp^2$-electron spin,
unlike in the UPS where two (i.e., a lone-pair singlet) electrons are localized.
Here again the bonds at ``a'' and ``X'' defining the edge of the vacancy are short.
The $\pi$ electron released from the vacancy delocalizes into the 2D network
around the vacancy in the PSPS. In fig.~\ref{charge-print} we show the distribution
of excess charge projected on the $x$-$y$ plane of the graphene sheet.

\begin{figure} 
\includegraphics*[width=8.0 cm, height=11 cm, angle=-90]{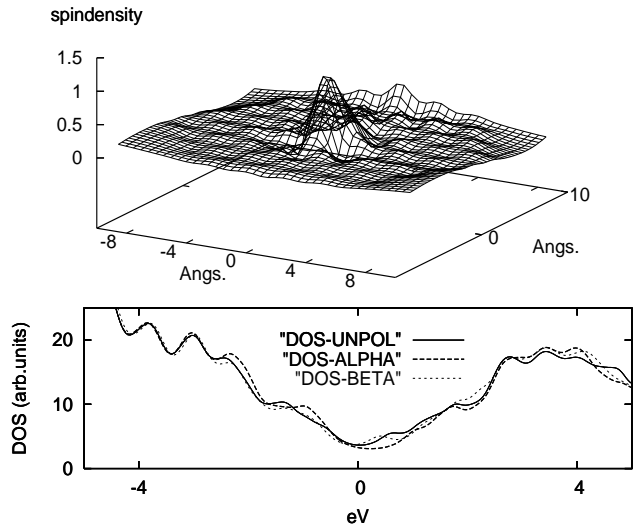}
\caption
{ The top panel gives the spin density ($n_\alpha-n_\beta$)
where $n_\sigma$ is the spin-dependent Mullikan charge on each C atom of
the planar structure.
The structure of the flat graphene piece is given in Fig.~\ref{mz-both2}.
The bottom panel shows the Electronic density of states (DOS) for the
 unpolarized system (bent structure) and
for the flat, polarized ground state. The $E_F$ is set to zero
and calculated with a Lorentzian broadening of 0.3 eV.
}   
\label{spinden}
\end{figure}

On comparing the panels (a) and (b) of Fig.~\ref{charge-print} it is
clear that the charge depletion caused by the unpolarized structure
is significantly larger than in the polarized structure of panel (b).
In the PSPS, (b), the $\pi$ electron released from the formation
of the vacancy re-distributes
itself  among the three pentagons.
This simplified discussion of the $\pi$-electron charge at the vacancy
has to be extended to allow the $\pi$-electron to spread in the full
 2D-network 
bounded by the edge-C atoms. The edge defines a
a ``zig-zag'' termination (saturated with hydrogen atoms in this calculation).
A plot of the Mullikan spin charge ($n_\alpha-n_\beta$),
 obtained from a spin-polarized DFT calculation
(using Gaussian-98)  is shown in Fig.~\ref{spinden} for the structure
of Fig.~\ref{mz-both2}. The total spin of the structure 
is consistent with a triplet state made up of the $sp^2$ electron on
the carbon atom `a' 
and the $\pi$-electron. It is also noted that some of the
spin-polarization resides at the graphene-edge seen
in the depth of the figure. This suggests that the finite-graphene
segment studied here is not large enough for the vacancy
to be isolated from the edge effects.
The electronic density of states (DOS) for these
are also shown in the bottom panel. 
The DOS of the unpolarized state was divided by 2 to compare
 with the polarized states, and a
Lorentzian broadening of 0.3 eV has been applied. The DOS shows substantial
filling around $E_F$, due to the localized states at $E_F$
created by the formation of the vacancy. 

 The finite-graphenes necessarily limit the $\pi$- electrons to a
 localized region. The boundary conditions imposed on the wavefunctions do not
 allow true extended states. This difficulty is overcome in calculations where
 periodic boundary conditions in the $x$-$y$ plane are used. We consider
 simulation cells with $N_a\times N_b$ graphene unit cells, each cell containing
 two carbon atoms, with a lattice parameter $a_0=2.47 $\AA.$\,$ and a C-C
 bond length of 1.42\AA$\,$ in unperturbed
 graphene. Calculations
 are mostly for $N_a=N_b$ for $ N_a$ =2,3,4, and 6, corresponding to
 $N=2N_a^2$ of 8, 18, 32
 and 72 carbon atom in the graphene simulation cell (SC). These calculations
  use the
 VASP\cite{vasp} plane-wave code (clearly, no terminating-H atoms are needed).
 Projected augmented-wave (PAW) 
pseudopotentials\cite{vasp} have been used for Carbon. The C pseudopotential
is well established and was used in several graphene-type calculations 
(e.g., Ref.~\onlinecite{jop,grp2,nieminen}). Removal of one C atom from the
test structures gives vacancy concentrations $x_v$ of 1/8, 1/18, 1/32, and 1/72.
The finite-graphene piece studied using Gaussian-98 calculations
contained 112 C atoms less one due to the vacancy. Of these, 46 were edge atoms. 
Thus, the finite graphene is possibly
 analogous to an extended system with a vacancy concentration $x_v$
of $\sim$ 1/66 {\it if} the edge effect is negligible in the above finite-system.
Similarly, the periodic simulation cell with 36 unit cells and 72 C atoms,
 $x_v$ = 1/72 is the most ``isolated'' vacancy studied here, the periodic image being
 six unit cells away. The systems with $x_v \ge 1/32$ increasingly contain vacancy-vacancy
interactions, since the periodic repetitions of the SC would place vacancies at  4, 3,
and 2 units of $a_0$ from each other. 

When vacancies are introduced into
graphene, the large stresses are relieved by
some neighbours (C-atoms near the vacancy) moving
towards the vacancy, while others move away. The distortion persists to at least the
third set of neighbours, and generates a bond-length distribution varying from
about 1.37\AA$\;$ to 1.45 \AA$\,$.
If several vacancies are present in the
system, the energy minimization enables some vacancies to migrate and form
divacancies. Thus the initial periodic structures can be ``annealed'' to
give structures which are (at least) at local energy minima. Here we
report results for systems with only one vacancy per simulation cell.
The structure optimization
involves (i) optimization of all atomic positions in a given simulation
cell with $N_a \times N_b$ unit cells and optimization of the 
cell parameter $a_0$, (ii)repeating spin-density functional
calculations till convergence of energies, and negligible
Hellman-Feynman forces (to less than 0.001 eV/Angstrom)
are obtained. Details of integration grids, simulation cells etc., are
similar to those discussed in previous work\cite{jop, grp2}.
Unlike in our results for finite graphene, no out-of-plane
distortions were found in the periodic systems studied here,
irrespective of the spin-polarization. The calculations for periodic
systems also reproduce a tri-pentagonal vacancy,
similar to the vacancy in finite graphenes. Both systems are shown in Fig.~\ref{mz-both2}. 

\begin{figure}
\includegraphics*[width=6.0 cm,height=7.8cm]{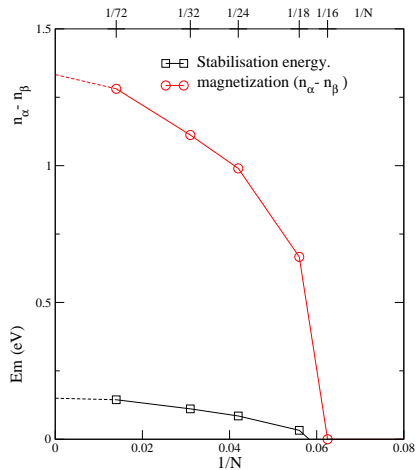}
\caption
{(Color online)The stabilization energy $E_{pol}-E_{unp}$, and the
spin polarization of the graphene vacancy, with one
vacancy per simulation cell (containing N lattice sites), and periodically
repeated. The polarized and unpolarized structures are planar.
The unpolarized planar ground state is stable for $x_v>\sim 0.06$.} 
\label{magfig}
\end{figure}

In Fig.~\ref{magfig} we summarize our results for the magnetism and stabilization
of graphene sheets containing  periodic realizations of vacancies. The unpolarized
 planar system is the ground state for systems with vacancies in excess of
 $x_v\sim 0.06$. At these higher vacancy levels, the atoms 
 move sufficiently close to each other and singlet pairing becomes possible.
 At lower $x_v$, the situation seems to be similar to the finite-graphene
 system where one of the $sp^2$ electrons
 remains unpaired, localized at the atomic
 site `a'.
 The uncoupled $\pi$-electron becomes
  a part of the 2D electron fluid since the C-skeleton remains planar.
  The intriguing result obtained from the VASP calculation is that
  the net magnetization (i.e., $n_\alpha -n_\beta$ ) accumulated 
  within the simulation
  cell can takes a fractional value in the spin-polarized case.
  This result remains robust when tested by increasing the size of the $k$-integration
  grids, and varying other technical parameters of the calculation. 
  One possibility suggests
  some sort of spin and charge fractionation. The shortened pair of bonds 
  at the atom `a', and at the atom `X' suggests that additional
  electron localization occurs around them.  The $\pi$-electron released
  in forming the vacancy
 remains near the vacancy, shared among the three pentagons 
 defined by the atoms `a', `b' and `c' 
 of the vacancy (Fig.~\ref{mz-both2}). However, a 1/3 spin-charge
  seems to be available for alignment with
  (or against) the spin already localized on the atom `a'.
   The calculations attempt to
  examine such possibilities 
  using explicit spin-density functional calculations which reliably
  predicts spin ground states. It is found that
  the spin localizes around the vacancy and aligns ferromagnetically for $x_v < 1/20$,
  (given the maximum spin of 1.333)
  and antiferromagnetically for $1/20 < x_v <1/16$, i.e., a magnetization
  of 0.66). Thus the low vacancy regime gives a
  net spin polarization $n_z=(n_\alpha-n_\beta)$ such that $1.333<n_z<0.666$,
  while the intermediate regime shows $0.666<n_z<0$. The high vacancy regime
  is unpolarized. The present calculation cannot sharply pin down the actual
  transitions or rigorously establish these suggestions. 
  
  The electronic DOS indicated in Fig.~\ref{spinden} for finite graphene 
  shows significant density of states near $E_F$. The situation is similar
  for infinite graphene with periodic vacancies. The high vacancy (unpolarized)
  systems have an energy gap. However, depending on the location of $E_F$ within
  the band structure, even within a rigid band model we see that
  the system may or may not be a metal, depending on the vacancy concentration $x_v$.
  This issue needs a fuller discussion in a separate study.
    
  The fore-going considerations suggest that the interaction of the 2D electrons
  containing the $\pi$-electron released from
  the vacancy, with the localized $sp^2$-electron spin on the atom 'a' has features of
  the Kondo problem, as well as fractional-charge/spin physics. Investigation of
  such matters would be best carried out within a more analytical approach, complemented
  by the results of computational approaches similar to the present study. 
    
   In conclusion, we have demonstrated, using finite as well as periodic structure calculations,
  that the ground state of a graphene sheet (held in free space at zero temperature) containing
  an isolated vacancy is spin polarized. When the concentration of
  vacancies (realized as a periodic
  array) exceeds $\sim 0.06$, the unpolarized structure becomes the ground state. All calculations
  using periodic boundary conditions lead to planar structures. Only the finite-graphene pieces
  which are unpolarized become distorted.


\begin{thebibliography}{99}
\bibitem{acronyms}
The 6-31G* basis and related acronyms are explained in
ref.~\cite{b3lyp}
\end{thebibliography}
\end{document}